\documentclass[twocolumn]{revtex4}
\usepackage{graphicx}
\usepackage{amssymb}

\def\E{{\rm E}_{10}}
\def\KE{{{\rm K}(\E})}

\def\cO{{\mathcal O}}

\def\11{{\mathbb 1}}

\def\keV{\,\rm keV}
\def\MeV{\,\rm MeV}
\def\GeV{\,\rm GeV}
\def\TeV{\,\rm TeV}

\def\MPL{M_{\rm Pl}}
\def\cd{\!\cdot\!}

\def\beq{\begin{equation}}
\def\eeq{\end{equation}}
\def\bea{\begin{eqnarray}}
\def\eea{\end{eqnarray}}

\begin{document}
\title{Planck Mass Charged Gravitino Dark Matter }
\author{Krzysztof A. Meissner$^1$ and Hermann Nicolai$^2$}
\address{
$^1$Faculty of Physics,
University of Warsaw\\
Pasteura 5, 02-093 Warsaw, Poland\\
$^2$Max-Planck-Institut f\"ur Gravitationsphysik
(Albert-Einstein-Institut)\\
M\"uhlenberg 1, D-14476 Potsdam, Germany
}

\vspace{3mm}

\begin{abstract} Following up on our earlier work predicting fractionally charged supermassive 
gravitinos, we explain their potential relevance as novel candidates for Dark Matter and   
discuss possible signatures and ways to detect them.
\end{abstract}
\maketitle

\vspace{5mm}

\section{Introduction.} 
In very recent work \cite{KM0} we have raised the possibility that, very unconventionally,  
dark matter (DM) could consist at least in part of an extremely dilute gas of very massive 
stable gravitinos, which are furthermore fractionally charged and possibly strongly interacting. 
In this article we wish to further investigate this possibility, and to discuss possible observable 
signatures and ways to search for them. A scenario based on such large mass DM candidates
is obviously very different from conventional models where the masses of putative 
DM constituents usually range from fractions of an eV (for axion-like DM) to the TeV scale 
(for WIMP-like DM); for supersymmetric DM candidates there is a particularly 
large variety of mass ranges, owing to the large number of different models.
With large mass a crucial issue is that of stability because superheavy particles 
participating in standard interactions can be expected to simply decay at a very 
early stage in the evolution of the Universe, unless a special mechanism 
is found that guarantees their survival to the present epoch. The crucial
new ingredient ensuring stability here is the fractional charge
of the DM candidates, together with their peculiar SU(3)$_c$ charge assignments, 
cf. (\ref{GravCharges}) below. We note that integrally charged DM candidates 
(`CHAMPs') have already been  discussed in the literature \cite{RGS}; likewise, 
and more exotically, DM candidates with very tiny (unquantized) charges have 
been considered \cite{BJRR,CK}. However, the latter proposals all  concerned 
sub-Planck mass particles. 

Although perhaps not so well known, there is already a substantial literature on the 
possibility of DM consisting of superheavy particles (SHDM), which is now receiving renewed 
attention in the light of the `no-show' of low energy supersymmetry at LHC and failed 
WIMP searches. Early work in this direction includes \cite{Markov,ACN,Sriv} where the 
DM constituents are assumed to be subject to gravitational interactions only. 
Later work incorporates inflationary cosmology into the picture \cite{SHDM1,SHDM2},
for instance by studying production of SHDM in the context of (large field) inflationary 
models; for more recent work, see also
\cite{PIDM,PIDM1,SHDM6} and references therein. A recurring feature of 
these studies is  that the SHDM particles are still assumed to have only weak 
(or even superweak) interactions with SM matter, whence they are commonly referred 
to as `WIMPZILLAs'. Since these considerations are mainly motivated by inflationary
cosmology, the mass of the DM constituents, though very large, is usually still assumed 
to be well below the Planck scale, but instead on the order of the scale of inflation
$\lesssim 10^{16}\,$ GeV \cite{SHDM1,SHDM2,PIDM,PIDM1,SHDM6}.
By contrast, the present model combines {\em Planck mass} with fractional electric charges 
and strong interactions of Standard Model (SM) type in a way that is completely new to the 
best of our knowledge~\footnote{In superstring inspired extensions of the SM the DM 
 is often part of a `hidden sector' where it may be subject to hidden strong 
 interactions while still having only weak interactions with SM matter. In that context, the notion
 `strongly interacting'  obviously does not refer to `ordinary' (SM) strong interactions.}, 
where, however, only the non-strongly interacting gravitinos would contribute 
significantly to DM. This is a main distinctive feature that sets the present  
proposal apart from earlier work on SHDM.

Perhaps even more importantly, the present scenario is based on a fundamental
ansatz that also aims for an explanation of the fermion content of the SM, and that
draws its inspiration from the huge duality symmetry E$_{10}$ that has been conjectured 
to underlie M theory \cite{DHN}.  More specifically,  and as explained in \cite{KM0}, 
our proposal relies on an attempt to embed the SM fermions into an M theoretic framework 
extending $N=8$ supergravity, which exploits the fact that after complete breaking 
of supersymmetry the remaining 48 spin-$\frac12$ fermions of this theory can be put 
in precise correspondence with the 3 $\!\times\!$ 16 quarks and leptons 
of the SM (including right-chiral neutrinos), following an insight originally 
due to \cite{GellMann}, see also \cite{NW}. We stress that the present version of
this proposal  does {\em not} necessarily require supersymmetry, 
but rather relies on $\KE$, an infinite-dimensional extension of the usual 
R symmetries of extended supergravities, and on the fact that the degrees
of freedom corresponding to a combination of eight 
massive gravitinos and 48 spin-$\frac12$ fermions at a given spatial point 
(obtained after appropriate decomposition of the $D\!=\! 11$ gravitino components)
constitute an irreducible unfaithful  spinorial representation of $\KE$ \cite{DKN,dBHP}. 
The unaccustomed feature here is that -- in contradistinction 
to accepted model building wisdom (as {\em e.g.} for GUT-type scenarios) -- 
the symmetry can be so enormously enlarged {\em without} increasing the 
size of the fermion multiplet. This interpretation hinges crucially on the assumption
of {\em emergent} spacetime in \cite{DHN} as there appears to be no way to achieve 
this in the framework of space-time based quantum field theory.

Apart from the intrinsic interest of incorporating infinite dimensional duality symmetries 
into unification in entirely novel ways, the main motivation for our proposal comes 
from the fact that, as far as the fermionic sector is concerned, it can make do with the 
particle content of the SM, that is, the observed three generations of quarks 
and leptons (including right-chiral neutrinos). This is in accord with indications from LHC 
that there may not be much in terms of new physics beyond the electroweak scale,  
and increasing evidence that the SM might survive up to the Planck scale more or
less {\em as is}, contrary to numerous still popular scenarios postulating a plethora of 
new particles at the TeV scale or just beyond. The possibility that the present framework 
also offers new options for DM is an extra incentive for further study.

\section{Main new features.}
Our proposal implies a number of highly unusual features for the DM gravitinos. We
caution readers that these features rely on a number of assumptions that are 
contingent on the proposal of \cite{DHN} according to which the full conjectured 
$\E$ symmetry and its compact subgroup $\KE$ manifest themselves only in a 
``near singularity limit" where space-time is assumed to de-emerge. 
With this reservation in mind let us list the special properties:

\begin{enumerate}
\item All gravitinos are assumed to be extremely massive with masses $m \sim \MPL$,
          or not too far from this scale. This high
          mass value is a consequence of the assumption that supersymmetry -- if at all present --
          is broken already at the Planck scale, leaving no room for low energy supersymmetry
          (with a single Majorana gravitino which would manifest itself in completely different 
          ways). In fact, as we already emphasized in \cite{KM0},  supersymmetry might 
          actually never be realized at {\em any} energy as a {\em bona fide} symmetry
          in the framework of space-time based quantum field theory.
\item The eight massive gravitinos split as 
         \beq\label{GravCharges}
         \left({\bf 3}\,,\,\frac13\right) \oplus \left(\bar{\bf 3}\,,\,-\frac13\right)
         \oplus \left({\bf 1}\,,\,\frac23\right) \oplus \left({\bf 1}\,,\, -\frac23\right)
         \eeq
          under SU(3)$\,\times\,$U(1). Identifying this SU(3) with SU(3)$_c$ as in
          \cite{KM0}, a complex triplet of gravitinos would thus be subject to strong 
          interactions (the alternative option of identifying SU(3) with the family symmetry 
          SU(3)$_f$ is disfavored for the reasons given in \cite{KM0}). Furthermore, 
          as explained in \cite{MN2}, the U(1) in (\ref{GravCharges}) is identified 
          with U(1)$_{em}$ whence all gravitinos carry {\em fractional electric charges}.
          As we will see, the SU(3)$_c$ assignments in (\ref{GravCharges}) lead to distinct 
          and well separated physical consequences: while the color singlets would mainly
          contribute to DM, the strongly interacting color triplet gravitinos could play a
          key role in explaining ultra-high energy cosmic rays (UHECR) \cite{MN3}.
\item Despite their strong and electromagnetic interactions with ordinary matter, the Planck 
         mass gravitinos would be stable. This is due to their fractional charges since there are simply no
         (confined or unconfined) fractionally charged final states in the SM into 
         which they could decay in a way compatible with SU(3)$_c\,\times\,$U(1)$_{em}$.
         Being stable all these particles should be around 
         us, though in extremely low abundance since the only processes changing the gravitino
         number are annihilations of gravitinos with anti-gravitinos and these are expected
         to be extremely rare over the whole history of the Universe after the Planck era, see below.
\item The color non-singlet gravitinos should form bound states with quarks so as to 
         avoid colored final states for temperatures $T < \Lambda_{QCD}$. Importantly, since 
         colored gravitinos have electromagnetic charge $+\frac13$ and anti-colored 
         ones $-\frac13$, with the known SU(3)$_c$ assignments of the 
         SM quarks there is no way to combine gravitinos with quarks or antiquarks 
         to build color singlet states that are neutral or integrally charged. 
         Of course, an important open question here concerns the strong interaction 
         dynamics of these superheavy 'meso-gravitinos' or 'baryo-gravitinos'. We here appeal to
         heavy quark theory (see {\em e.g.} \cite{MM}), where the confinement scale
         is set by the difference between the mass of the bound state meson and 
         the mass of the heavier constituent, which is usually of the 
         order of $\Lambda_{QCD}$.
\item Independently of whether they are in bound states or not,  the gravitinos do not
         interact in any way with the CMB despite their electric charges. This follows 
         immediately from the Thomson formula, according to which the total cross 
         section (for low energy photons) is proportional to the square 
         of Compton wave length of the scatterer (see {\em e.g.} \cite{Nachtmann}).
         In our case, the relevant scale is the Planck length, hence 
         the cross section is suppressed by a huge factor, and thus completely negligible.
\item By contrast, interactions with charged non-strongly interacting matter are governed by the Rutherford  
         formula, and therefore much like ordinary charged particle interactions.
         Being essentially at rest w.r.t. the cosmic frame (see below) 
         the gravitinos would merely `stir' the surrounding charged light matter 
         particles but not affect their thermalization. Unlike common 
         DM candidates with masses $\lesssim \cO(1 \TeV)$ they would thus not 
         produce any significant dissipative effects in the evolution of the universe,
         except possibly in the very earliest moments after the Planck era.         
\item The color singlet gravitinos in (\ref{GravCharges}) (which do not participate
         in strong interactions) are never in thermal equilibrium during the 
         evolution of the Universe after the Planck era, so common astrophysical 
         wisdom (see e.g. \cite{KT,Masiero}) does not apply. This can be seen as 
         follows. The inverse collision time is given by the standard formula
         $\Gamma \sim \langle n\sigma v \rangle$  where $n$ is the particle
         number density and $\sigma$ the annihilation cross section. For the 
         annihilation of a charged gravitino-antigravitino pair of mass $M$ and charge $e$
         into a pair of spin-$\frac12$ fermions the  cross section behaves as 
         $\sigma v \propto \alpha^2/M^2$, where $v$ is the velocity of the incoming particles 
         (which is small) and $\alpha \equiv \alpha(M)=e(M)^2/4\pi$. Thermal 
         equilibrium requires $\Gamma \gtrsim H \sim T^2/\MPL$ \cite{KT}, so with the general
         formula $n=g(\mu T/(2\pi))^{3/2} e^{-\mu /T}$ for particles of mass $\mu$ ($g=4$ for each 
         massive gravitino species) this constraint translates into an approximate condition
         \beq
         \left(\frac{\mu}{T}\right)^{\frac12} e^{-\mu/T} \,\gtrsim \, \frac{\mu}{\alpha(\mu)^2\MPL }
         \eeq
         which for $\mu\sim \MPL$ and $T<\MPL$ can never be satisfied (note that 
         $\alpha(\mu)\lesssim O(0.1)$ for all SM gauge couplings over the whole range of $\mu$
         from the weak scale up to $\MPL$). In other words, the non-strongly interacting Planck  
         mass DM gravitinos would be frozen out from the very beginning. Their abundance 
         thus cannot be estimated from thermal equilibrium but requires a ``pre-Planckian"    
         explanation. The rapid decrease of the annihilation cross section $\propto \MPL^{-2}$
         of color singlet gravitinos, together with their extreme dilution, also shows that they 
         have no effect on the CMB or UHECR processes.
\item By contrast, the strongly interacting (color triplet) gravitinos can reach thermal equilibrium, 
         due to their strong interactions and the fact that the relevant cross section
         does {\em not} decrease with energy \cite{Wibig}, unlike for the color singlet gravitinos,
         thus allowing for an estimate of the color triplet gravitino density. This
         density turns out to be too small for the strongly interacting gravitinos to contribute in 
         any significant way to DM, unlike the color singlet gravitinos. However, they could
         play a key role in explaining UHECR events \cite{MN3}.           
\item If any signals originating from DM gravitinos were to be found they would provide direct 
         access to Planck scale physics. We also note that for a Planck mass particle the Compton
         wavelength coincides with the Schwarzschild radius (both are thus nearly equal to
         the Planck length). When viewed as mini black holes our gravitinos are very close
         to, but strictly below extremality (because $\frac23 < 1\,$!). As a result the attractive 
         force for oppositely charged gravitinos is almost doubled, while for charges of the 
         same sign we are very close to a `force-free' BPS-type situation. Possible 
         consequences of this fact for cosmological issues (such as structure formation)
         remain to be explored, however.        
\end{enumerate}

Let us emphasize that the present scenario is completely different from earlier proposals 
with {\em light} neutral (Majorana) gravitinos as DM candidates. In conventional scenarios of 
low energy supersymmetry and supergravity, gravitinos do {\em not} carry SM charges (this would 
require $N$-extended supergravities with at least $N\geq 2$, but for these one cannot 
have chiral gauge interactions with non-composite gauge bosons). Depending on their 
mass, such neutral gravitinos would either decay into 
lighter supersymmetric particles (neutralinos) via the Noether interaction present 
in any supergravity Lagrangian, or themselves contribute to DM if they cannot
decay. Either of these scenarios differs from the present one since our gravitino DM 
candidates {\em do} participate in SM interactions, but cannot decay because of the 
absence of suitable fractionally charged final states in the SM, despite their interactions 
with SM matter. So it is precisely the exotic gravitino charge assignments that can 
make our Planck mass gravitinos survive to the present epoch.

\section{Some properties of superheavy gravitino dark matter}
As we said, the possibility of DM carrying SM charges  \cite{RGS} 
has already been considered in the literature, although not very prominently because 
DM is usually assumed to interact only very weakly with SM matter, apart from 
their gravitational interactions \cite{KT}. The relevant analyses are obviously 
very model dependent, see e.g. \cite{RGS}, and usually apply only for much lighter 
DM constituents, so accepted cosmological bounds may be invalid for the case 
of masses of the order of $\MPL$ considered here.
In fact, at least in more conventional DM scenarios, electrically charged DM is already 
very strongly constrained by existing data: it is either completely diluted, or otherwise 
the electric charges of putative DM particles must be extremely small. Indeed, the most 
stringent cosmological bound on the charge of DM particles of mass $m$ 
is \cite{milli1,milli2,NNP}
\beq
|q|\;\lesssim \; 7.6\cdot 10^{-10}\left(\frac{m}{1\TeV}\right)^\frac12
\eeq
with 90\% confidence limit. 
For the DM candidates usually discussed (axion-like or WIMP-like, or any kind of new particle
associated with low energy supersymmetry) which are assumed to have  masses 
$\lesssim\, \cO(1\TeV)$ this implies that the allowed charges are $\lesssim \cO(10^{-10})$. 
This completely excludes charged DM of any conventional type (a possible way out here 
would be to invoke new U(1) gauge interactions but there is neither observational 
evidence nor any compelling theoretical reason for them).
Remarkably, however, if we assume the DM particle to have Planck scale mass, then
the admissible charge comes out to be of order unity: for $m \sim 10^{19}$ GeV  
the above formula gives
\beq\label{charge}
|q|\;\lesssim \;7.6\cdot 10^{-2}
\eeq
Taking into account the theoretical uncertainties and model dependencies, this value is quite
compatible with charges of order one!

The assumed large mass of the DM gravitinos has another important 
consequence: it is a well known result in general relativity that in the 
course of the expansion of the Universe any peculiar motion w.r.t. the cosmic frame
of a massive particle out of equilibrium decreases as the inverse of the scale factor.
So whatever the initial velocity distribution was shortly after the Planck era,
it is reduced by a factor of $a_{\rm PL}$/$a_{now} \sim 10^{-30}$, despite 
occasional scatterings with particles that will not appreciably change 
the energy because of the large mass. In other words, the superheavy gravitinos 
would be effectively at rest w.r.t. the CMB rest frame, with a very small velocity 
dispersion. However, the situation changes when structures are formed:
then the heavy gravitinos can be trapped by a galaxy and subsequently move 
along geodesics, with a velocity of the order of several hundred km/s relative 
to the CMB ({\em i.e.} the escape velocity of the Milky Way galaxy at a typical 
distance from the center). Not much appears to be known about the motion of trapped 
DM  {\em relative to} luminous matter inside galaxies, but it seems reasonable to assume 
that it simply `moves along' with luminous matter, with a small velocity dispersion.


In conclusion we would expect our DM candidates to move with an effective velocity 
of some tens of kilometers per second w.r.t. Earth (this follows also from
from simple considerations based on the virial theorem in Newtonian physics).
In this case their non-relativistic kinetic energy is of order
\beq
E\, \sim \, \frac12 \MPL v^2 \, \sim \, 10^{20}\ {\rm eV}
\eeq
To estimate their penetration depth we recall that a proton of velocity 400 km/s 
(i.e. with kinetic energy $\sim 1\keV$) in iron loses approximately $300\,\MeV$ per
centimeter \cite{NIST}. Being subject to similar electromagnetic interactions this implies about 
ten times smaller energy loss rate for our DM candidates (because of the charge squared 
factor $\frac49$ or $\frac19$), so their their range would be
\beq
R\,\sim \, \frac{E}{30\ {\rm MeV/cm}}\,\sim \, 3\cdot 10^{10}\ {\rm m}
\eeq
Consequently, these particles will easily pass through the Earth without appreciable 
change in energy. Nevertheless, because of their electromagnetic interactions they will 
uniformly ionize their surroundings, leaving a straight ionized track all along
their path. This track would have a lateral extension of a few nanometers, and would
thus not be visible in ordinary light. By contrast, the passage through the sun
or some other star might lead to some absorption, due to the much larger stopping power
of a plasma environment. But even assuming that all gravitinos hitting the sun were stopped, and
taking into account their low abundance and flux rates (see below), the total amount 
of gravitino DM captured inside the sun would be rather tiny and
therefore not affect stellar processes in any significant way (neutron star evolution 
also allows for Planck mass DM since known bounds only exclude 
masses $<10^{16}$ GeV  \cite{G}).

To estimate the flux, we recall that the mass density of DM in our galaxy 
in the proximity of the Solar System is usually given as \cite{WdB}
\beq\label{DM}
\rho_{DM} \,\sim \,0.3\cdot 10^6 \GeV\cd m^{-3}  \,.
\eeq 
If DM is made out of Planck mass particles, this means 
roughly $3\cdot 10^{-14}$ particles per cubic meter, that is, a very low abundance 
to compensate for the very large DM constituent mass.
Putting in an estimated average velocity $\beta \sim 10^{-4}$ (that is, on the order
of the Earth's orbital velocity around the Sun) we arrive at a flux estimate of
\beq\label{FLUX}
\Phi \,\lesssim\,   10^{-14}\ {\rm cm}^{-2}{\rm s}^{-1}{\rm sr}^{-1}\,
\sim \,0.003\ {\rm m}^{-2}{\rm yr}^{-1}{\rm sr}^{-1}
\eeq
We stress that apart from uncertainties about the velocity distribution, there
are also uncertainties about the assumed DM density in our vicinity which could 
be subject to potentially large local variations ---  the comparison of this value 
with the experimental bounds may provide a hint on the scale of these variations.

\section{Prospects for detection}

What are the prospects for actually detecting such non-relativistic 
superheavy DM candidates? Searches for ionized tracks coming from DM particles 
have a long history, see {\em e.g.} \cite{PM} and references therein. Currently
there are several direct WIMP searches (see {\em e.g.} \cite{CG,PM}), as
well as  accelerator experiments (for example MoEDAL at CERN \cite{moedal}). 
Alternatively one might consider an underground {\em paleo-detector} that could 
identify long ionized tracks preserved in rock, and discriminate them from muon tracks or 
neutrino induced events. In fact, there are projects to detect tracks in old rocks \cite{AD} 
and plans for experiments to look for superheavy DM with multiple scatterings \cite{BBLR}. 
There are also limits on the allowed fluxes for fractionally charged superheavy 
DM \cite{Ambrosio}, but these concern only {\em ultra-relativistic} particles ($\beta > 0.25$). 

Nevertheless, it appears that WIMP-like searches of the type currently pursued
are unsuitable for detecting superheavy gravitinos. For the existing
experiments LUX \cite{LUX}, XENON1T \cite{XENON} and DAMA/LIBRA \cite{DAMA},
the flux (\ref{FLUX}) is way too low to be seen, as their fiducial volumes/fiducial 
areas are simply too small for a detection. For instance, the LUX experiment had 
250 kg of liquid Xenon (with density 2.9 g/cm$^3$ it gives a box of effective 
volume (44 cm)$^3$) and was effectively operating for 95+332 days. 
With the estimated flux (\ref{FLUX}) this would give a total of 
$0.003 \times 4\pi \times 0.44 \times 0.44 \times 427/365 \sim 0.009$  hits over the whole 
time of exposure; the estimates for XENON 1T are similar. DAMA/LIBRA had comparable
fiducial area  but a much longer exposure; however, it looked only for single hit events
very different from our putative gravitino tracks. 
Likewise large detectors used in accelerator experiments 
have the triggering procedures focused only  on relativistic particles 
(as for CMS, ATLAS or Superkamiokande). 

Much more relevant to the present proposal are past searches for magnetic 
monopoles with very large fiducial area/volume  that were conducted up until 2000, 
and that have already established significant limits, see \cite{MMreview} for an early review, 
and the MACRO report \cite{MACRO} for a final summary (though magnetic monopoles 
were never considered to be serious candidates for explaining DM). Indeed, GUT mass 
magnetic monopoles would produce signals very similar to the ones postulated here
(although the degree of ionization caused by monopoles might be somewhat different).
However, because of their large magnetic charge their velocity dispersion is expected 
to be much larger than in our case, as magnetic monopoles can be accelerated to relativistic 
speeds by galactic magnetic fields. It was presumably for this reason that past 
searches were limited to velocities in the range $10^{-4} < \beta < 1$ \cite{MMreview}. 
By contrast, we expect gravitinos to have velocities of the order of $\beta \sim 10^{-4}$, 
and then gravitino induced signatures would have no natural background 
(except possibly extremely heavy magnetic monopoles). Hence the cleanest 
way to search for superheavy charged gravitinos seems to be a dedicated time-of-flight 
underground experiment looking for {\em slow} ionizing particles. This could be done 
for instance by resuscitating and/or redesigning the old experiments to cover the so far
little explored velocity range $10^{-5} < \beta < 10^{-4}$ aiming at lower fluxes
than previously~\footnote{Ref. \cite{MACRO} mentions
 40 events that were subsequently discarded as spurious. In the light of the present
 proposal it is perhaps worthwhile to have another look, as we do not know to what 
 extent the exclusion of these events depended on the monopole hypothesis.}.

One may note that from the present perspective the negative results of MACRO 
\cite{MACRO} could also be interpreted as evidence for a significantly lower DM density 
in the vicinity of the Earth than the usually quoted average value of $0.3 \GeV\cd\ cm^{-3}$.
The comparison of the expected flux given in (\ref{FLUX}) with the MACRO bound 
\cite{MACRO} for velocities $\beta=10^{-4}$ and fluxes  
$\Phi \sim 3 \cdot 10^{-16} \ {\rm cm}^{-2}{\rm s}^{-1}{\rm sr}^{-1}$ could be a hint
that the actual value of the local DM density  may be significantly lower than the usually 
assumed value (\ref{DM}). Indeed, in \cite{MN3} we have put forward the hypothesis
that a large fraction of the DM in galaxies could reside inside stars, thereby depleting the DM
content of interplanetary space. In that case the only remaining option for detecting
DM might be a paleo-detector with very long exposure time, similar to  \cite{AD}.

We finally note that, while posing a considerable challenge, experiments 
searching for exotic DM candidates with properties and fluxes similar to the 
ones predicted by the present scheme are of interest in their own right, 
independently of the theoretical motivation given here.\\

\noindent
 {\bf Acknowledgments:} We thank Miko{\l}aj Misiak and S{\l}awomir Wronka for discussions.
 K.A.~Meissner thanks AEI for hospitality and support; he was 
 partially supported by the Polish National Science Center grant DEC-2017/25/B/ST2/00165.
 The work of  H.~Nicolai has received funding from the European Research 
 Council (ERC) under the  European Union's Horizon 2020 research and 
 innovation programme (grant agreement No 740209).

\end{document}